\documentclass{article}

\usepackage{arxiv}

\usepackage[utf8]{inputenc} % allow utf-8 input
\usepackage[T1]{fontenc}    % use 8-bit T1 fonts
\usepackage{hyperref}       % hyperlinks
\usepackage{url}            % simple URL typesetting
\usepackage{booktabs}       % professional-quality tables
\usepackage{amsfonts}       % blackboard math symbols
\usepackage{nicefrac}       % compact symbols for 1/2, etc.
\usepackage{microtype}      % microtypography
\usepackage{lipsum}		    % Can be removed after putting your text content
\usepackage{graphicx}
\usepackage[numbers,square]{natbib}
\usepackage{doi}
\usepackage{amsmath}
\usepackage[bottom]{footmisc}
\usepackage{multirow}
\usepackage{footnote}
\hypersetup{hidelinks}

\title{DEPAS: De-novo Pathology Semantic Masks using a Generative Model}

\date{}


\author{ \href{https://orcid.org/0000-0002-5006-9300}{\includegraphics[scale=0.06]{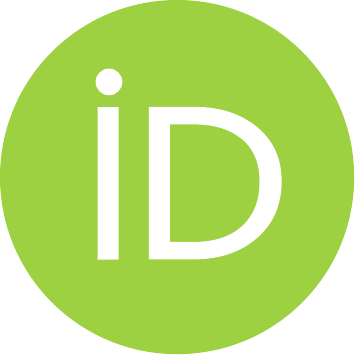}\hspace{1mm}Ariel Larey$^{1,2}$}\\
    Technion IIT\\
	    \And
	\href{https://orcid.org/0000-0002-0939-3379}{\includegraphics[scale=0.06]{orcid.pdf}\hspace{1mm}Nati Daniel$^{1}$} \\
    Technion IIT\\
		\And
	\href{https://orcid.org/0000-0003-4591-7018}{\includegraphics[scale=0.06]{orcid.pdf}\hspace{1mm}Eliel Aknin$^{1,3}$} \\
	Technion IIT\\
		\And
	\href{https://orcid.org/0000-0003-3304-4854}{\includegraphics[scale=0.06]{orcid.pdf}\hspace{1mm}Yael Fisher$^{4}$} \\
	Rambam Health Care Campus\\
		\And
	\href{https://orcid.org/0000-0002-5345-8491}{\includegraphics[scale=0.06]{orcid.pdf}\hspace{1mm}Yonatan Savir$^{1,}$\thanks{Corresponding author, e-mail: yoni.savir@technion.ac.il. $^{1}$Department of Physiology, Biophysics and System Biology, Faculty of Medicine, Technion Israel Institute of Technology, Haifa, Israel.
    $^{2}$Faculty of Computer Science, Technion Israel Institute of Technology, Haifa, Israel. 
    $^{3}$Faculty of Industrial Engineering, Technion Israel Institute of Technology, Haifa, Israel.
    $^{4}$Division of Pathology, Rambam Health Care Campus, Haifa, Israel.}} \\
	Technion IIT\\
}

\renewcommand{\headeright}{Larey et al.}
\renewcommand{\shorttitle}{DEPAS: De-novo Pathology Semantic Masks using a Generative Model}

\begin{document}

\maketitle

\begin{abstract}
The integration of artificial intelligence into digital pathology has the potential to automate and improve various tasks, such as image analysis and diagnostic decision-making. Yet, the inherent variability of tissues, together with the need for image labeling, lead to biased datasets that limit the generalizability of algorithms trained on them. One of the emerging solutions for this challenge is synthetic histological images. However, debiasing real datasets require not only generating photorealistic images but also the ability to control the features within them. A common approach is to use generative methods that perform image translation between semantic masks that reflect prior knowledge of the tissue and a histological image. However, unlike other image domains, the complex structure of the tissue prevents a simple creation of histology semantic masks that are required as input to the image translation model, while semantic masks extracted from real images reduce the process's scalability.
In this work, we introduce a scalable generative model, coined as DEPAS, that captures tissue structure and generates high-resolution semantic masks with state-of-the-art quality. We demonstrate the ability of DEPAS to generate realistic semantic maps of tissue for three types of organs: skin, prostate, and lung. Moreover, we show that these masks can be processed using a generative image translation model to produce photorealistic histology images of two types of cancer with two different types of staining techniques. Finally, we harness DEPAS to generate multi-label semantic masks that capture different cell types distributions and use them to produce histological images with on-demand cellular features. Overall, our work provides a state-of-the-art solution for the challenging task of generating synthetic histological images while controlling their semantic information in a scalable way.

\end{abstract}

% keywords can be removed

\section{INTRODUCTION}
As the adoption of digitized histopathologic slide images became widespread, the use of Artificial Intelligence (AI) methods in the digital pathology field increased. In particular, computer vision and deep learning methods are used to automate and improve various tasks such as image analysis, diagnostic decision-making, and disease monitoring \citep{czyzewski2021machine, daniel2022deep, larey2022fron}. 

However, data limitations pose a major challenge in digital pathology and include issues related to data scarcity, variability, privacy, annotation, bias, quality, and labeling. Data scarcity and variability can make it difficult to train and evaluate computer algorithms for digital pathology, as there may not be enough data available for certain decision thresholds.  Data bias is another concern, as digital pathology datasets may be biased toward certain populations, which can limit the generalizability of algorithms trained on them. Data labeling can also be subjective and dependent on the expertise of the labeler leading to inaccuracies. An emerging solution to these challenges is generating synthetic images.

The field of generating synthetic images became more popular during the last years after the Generative Adversarial Network (GAN) \citep{goodfellow2020generative} was introduced. In this approach, a ‘discriminator’ model is designed to discriminate between real data to fake data. A different model coined ‘generator’ is trained to produce synthetic data that will be plugged into the ‘discriminator’ during training. On the one hand, the ‘generator’ is trained in a way where the discriminator doesn’t distinguish between the real data and the generated synthetic data. On the other hand, the discriminator model is trained to discern between the two correctly. It means the generated data is challenging the discriminator to get the best results.

In the classic approach of image generation (coined Vanilla GAN), the input for the generator is sampled from a given distribution, and then is processed into a synthetic image. More advanced techniques called Conditional GAN (C-GAN) \citep{mirza2014conditional} supply information about the required type of generated data and plug it into the different GAN models, to control the type of generated data. Some approaches such as pix2pix \citep{isola2017image} took this technique further and supply the generator with more detailed information at the pixel level. In this approach, the generator receives a semantic label mask as an input, and each pixel is generated to belong to its corresponding label from the given semantic mask. This image translation approach has the advantage of yielding pairs of images and semantic labels, that could be used in different tasks that require these pairs (e.g. segmentation), unlike the classic approach where the synthetic images lack semantic information. Yet, in some cases, the scarcity of semantic masks will be caused due to their creation complexity.

A special case is the generation of synthetic histology images, where the semantic masks consist of various tissue types and complicated patterns resulting from the complex nature of the tissue. A naïve solution uses tissue masks extracted from real histology images in the image translation pipeline, but the dependency on limited real images during the generation process causes a limited number of semantic masks. Thus, in this case, image translation models will not be scalable since the semantic masks have an integral part in their pipeline, while Vanilla GANs are scalable due to their only dependency on the scalable sampling process.

\begin{figure}[thpb]
\centering
	\includegraphics[scale=0.6]{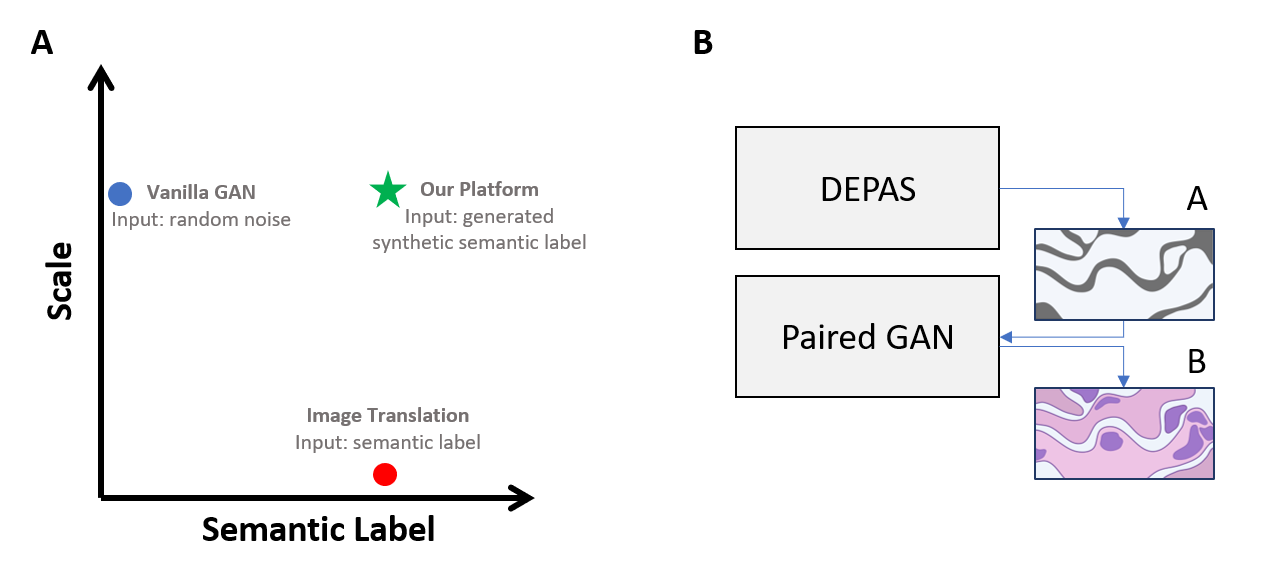}
	\caption{(A) Illustration of the tradeoff between Image Translation GANs to the Vanilla GANs. The former generates synthetic images based on their semantic labels. In this case, the scalability is bounded when the quantity of semantic labels is limited. On the Other hand, Vanilla GANs lack semantic information but can produce an unlimited number of synthetic images. (B) Our platform resolves this challenge in the histology domain with a dual-phase generative system. The first step includes generating semantic masks of tissue labels using a novel architecture of a Vanilla GAN coined DEPAS. Then, the generated masks are processed by a paired image translation GAN (such as pix2pixHD) to produce the synthetic histology RGB image.}
	\label{f:f1}
\end{figure}

In this study, we show how our dual-phase pipeline overcomes the tradeoff (Fig.~\ref{f:f1}), and generates pairs of histology synthetic semantic masks and images in a scalable design. We introduce DEPAS, a generative model that captures tissue structure and generates high-resolution semantic masks with state-of-the-art quality for three different organs: skin, lung, and prostate. Moreover, we also show that these masks can be processed by pix2pixHD \citep{wang2018high}, a generative image translation model that supports high-resolution images, to produce photorealistic RGB tissue images (Fig.~\ref{f:f1}). We demonstrate it for two types of staining: H\&E and immunohistochemistry. This pipeline, on the one hand, generates pairs of semantic masks and histology images, and on the other hand, is scalable since it does not require real masks during inference.

\section{RELATED WORK}
\subsection{Medical Synthetic Images}
The use of generative models to produce synthetic images was explored in numerous works in the medical field. Image translation frameworks are widely used, such as models that generate endoscopy images given binary semantic masks \citep{adjei2020gan}, transform between radiological images \citep{armanious2020medgan}, and convert between histology staining types \citep{lysik2022he}. Another image translation work is DeepLIIF \citep{ghahremani2022deepliif}, which provides for a given IHC image several outputs including stain deconvolution, segmentation masks, and different marker images. Other types of generative frameworks are common as well. DCGAN framework generates synthetic images from a sampled noise input and processes it through convolution-layers architecture. It was used in several applications that generate medical images such as MR images \citep{divya2022medical}, eye diseases images \citep{smaida2021dcgan}, X-ray images \citep{puttagunta2022novel}, and breast cancer histological images \citep{blanco2021medical}. PathologyGAN \citep{quiros2019pathologygan} introduced a novel framework that generates high-quality pathology images in the size of 224X224 pixels. \citep{guibas2017synthetic} introduced a two-step pipeline that is similar to ours. In the first step, they generate binary vessel segmentation masks using DCGAN, next they generate the RGB retinal image. Their pipeline provides synthetic images in the size of 512X512 pixels size. However, our pipeline provides higher-resolution (x2) synthetic images in the challenging field of histology. We focus on the first phase of learning the complex geometry structure that is reflected in the semantic label mask. We show that DCGAN is not sufficient for this task and introduce DEPAS as an improved architecture to overcome the challenges in the high-resolution histology domain.

\subsection{Discrete Predictions}
The first step of our pipeline requires predicting discrete semantic masks. In this work, we focus on the binary scenario where there are two labels in the semantic mask – tissue and air. However, the binary output of the generator should be obtained by a step-function, but this non-differentiable operation can break the backpropagation of the optimization objective's gradients through the discriminator to the generator. A reasonable solution is by replacing the discrete output operations with continuous relaxations such as Sigmoid during training, and applying the discrete operation only during the test \citep{neff2017generative}.  \citep{bengio2013estimating} proposed to use binary neurons in machine learning models via straight-threw-estimators, where the binary operator is applied during training in the forward pass, but is ignored and treated as an identity function in the backward pass. \citep{dong2018training} explored the generative use of a deterministic binary neuron and stochastic binary neuron and introduced the BinaryGAN. We investigated the different approaches and found that in the high-resolution histology domain, the best performance was achieved by the Annealing-Sigmoid. In this approach, the last layer of DEPAS’s generator is a Sigmoid whose slope is increased gradually during training toward the step-function \citep{chung2016hierarchical}. 

\begin{figure}[t]
\centering
        \includegraphics[ width=\textwidth]{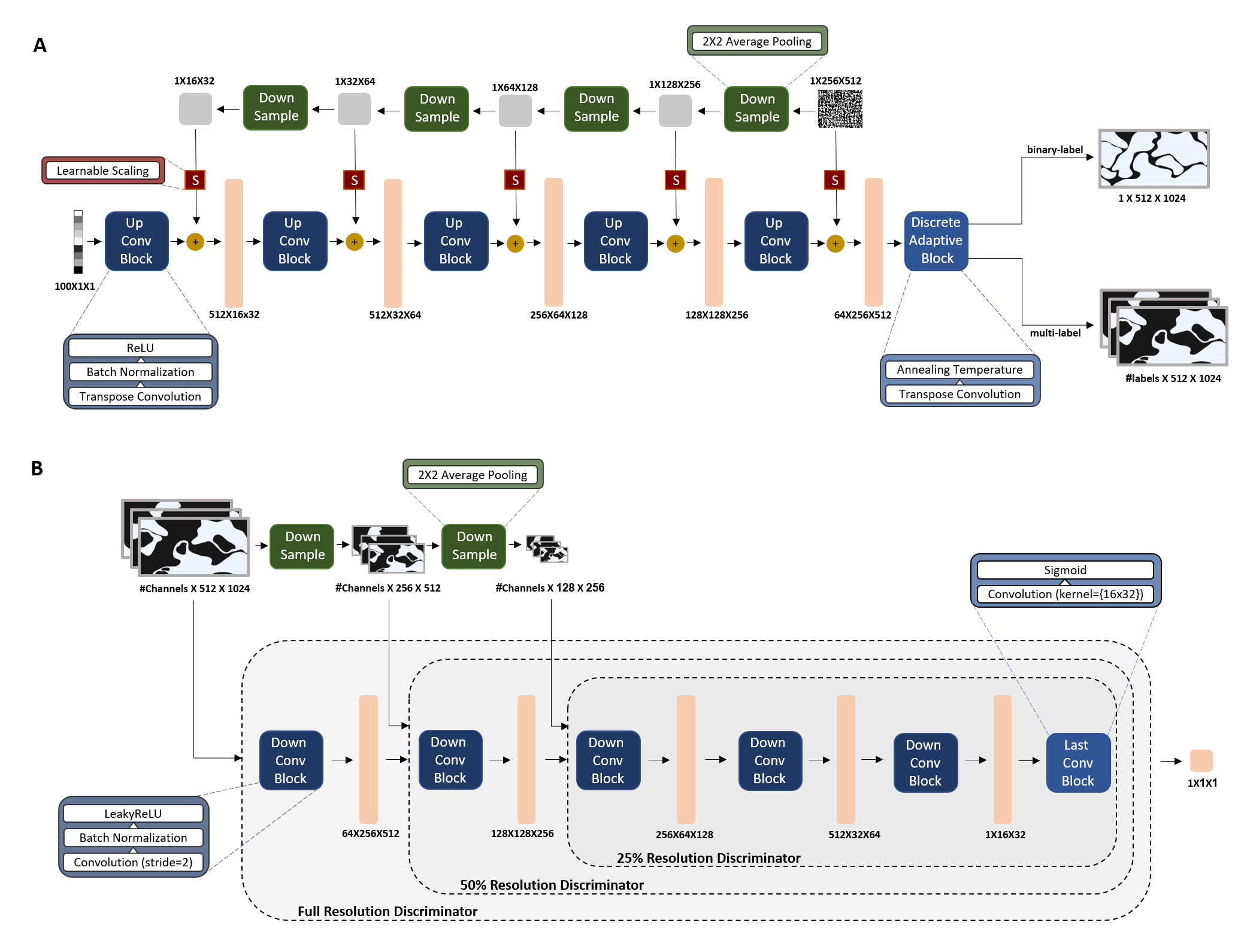}
	\caption{Architecture of DEPAS. (A) The generator decodes semantic masks from latent noise. It consists of five transpose convolution layers where each one of them followed by Batch Normalization and ReLU activation. Another element of stochasticity is added to the hidden layers in the spatial dimension after being scaled. Finally, the last feature-maps are processed by the Discrete Adaptive block that outputs a semantic mask in the two-label case, or multiple masks in the multi-label case. (B) For training, we use three discriminators that support different resolutions of images. Each one of them encodes the corresponding image into a scalar which represents the probability that the image is real. The encoding is processed by convolution layers where each one of them is followed by Batch-Normalization and LeakyReLU activation.}
	\label{f:f2}
\end{figure}

\section{METHODS}
Our pipeline includes two main phases. The main focus of this study is on the first step where we learn the internal geometry of the digitalized histology tissue. For this task, we designed a generative architecture coined DEPAS that captures the tissue’s morphology and expresses it by a semantic mask. The second phase is an image translation task to transfer the discrete semantic mask to an RGB photorealistic image of the tissue.

\subsection{DEPAS Architecture}
To enhance scalability, the generative process of producing synthetic tissue masks is initialized by sampling noise from a given distribution and applying it to the model (Vanilla GAN). The mechanism is based on the DCGAN architecture that was used by \citep{guibas2017synthetic} and consists of multiple convolution blocks in its generator and discriminator. In DEPAS we adjusted the DCGAN layers to the high-resolution size of 512X1024 pixels output and included three main extensions.

(1) Discrete Adaptive Block. In our case, where the discriminator should obtain a binary mask during training, we require a binary output from the generator where every pixel indicates one of the two classes – Tissue or Air. Thus, we replaced the DCGAN's last block with this module (Fig.~\ref{f:f2}A). Instead of the non-differential step function, we use a Sigmoid activation with a high slope to mimic the former and yield a pseudo-binary differential output. For optimal convergence, we initiate the Sigmoid with its base slope of $1$ and increase it gradually during training (Annealing-Sigmoid, AKA AS). That is, in every iteration, the generator produces a Bernoulli probability that becomes more deterministic during training in differentiating the two classes. Formally, at iteration $t$, the AS is:
\begin{equation}
{AS}_{t} = \frac{1}{1+e^{-\delta_{t}*x}} \label{eq_sig}
\end{equation}

Where $x$ is the input for the element-wise operation, and $\delta_{t}$ determines the Sigmoid’s slope at iteration $t$. To increase the slope, we require that $\delta_{t+1}>\delta_{t}$, and for initialization with the basic Sigmoid, we define $\delta_{t=0}=1$. 
Furthermore, we extend this approach to cases where there are more than two labels in the desired semantic mask. For example, in the case where the tissue itself has several types of morphology (e.g. tumor tissue, non-tumor tissue, and air) we will use the multi-label approach. In this scenario, we generalize the binary distribution to the multinomial distribution by designing the ‘Discrete adaptive Block’ to produce a multi-channel feature map, where each channel represents a different class. The feature maps are then applied to an Annealing-Softmax-Temperature (AST) activation. Instead of the non-differential Argmax function, we use a channel-wise Softmax layer with a low temperature to mimic a deterministic decision of the generated class for each pixel. Similarly to the binary situation, we initiate the Softmax temperature with its base value of $1$, and decrease it gradually during training. I.e. every iteration, the generator produces for each pixel its classes probabilities that become more deterministic during training.  Formally, at iteration $t$, the probability for class $c$ provided by the AST is:
\begin{equation}
{AST}_{t,c} = \frac{e^{\frac{x_{c}}{T_{t}}}}{\sum_{j}^{} e^{\frac{x_{j}}{T_{t}}}} \label{eq_temp}
\end{equation}

Where $x_i$ is the input for the element-wise operation of the channel that corresponds to class $i$, and $T_{t}$ determines the Softmax’s temperature at iteration $t$. To increase determinism, we require that $T_{t+1}<T_{t}$, and for initialization with the standard Softmax, we define $T_{t=0}=1$. 
Both binary and multi-label scenarios consist of the ‘step-function’ and ‘argmax’ operations respectively for inference. However, for training, where gradients should backpropagate through these layers, the non-differential operations are replaced by differential operations that adapt the former’s attributes gradually. 

(2) Spatial Noise. In the standard DCGAN’s implementation, latent vectors $z$ are drawn from a Gaussian distribution as the input for the generator. The sampling is performed channel-wise. That is, a sampled input noise is a one-dimensional latent vector where each element represents an initial channel with a spatial size of 1X1 pixels without any noise diversity in the spatial domain. When the feature maps are spatially increasing in the forward pass (via the transpose convolution layers), they are prone to become repetitive in the spatial aspect. In our case, where DEPAS provides high-resolution semantic masks, this phenomenon is significant. We address it by adding noise in the spatial domain as well. We draw a 2D array from a Gaussian distribution and inject it spatially into the hidden layers of the generator after resolution and scale adjustments (Fig.~\ref{f:f2}A).

(3) Multi-Scale-Discriminators. Our pipeline generates high-resolution synthetic images in the size of 512X1024 pixels. For comparison,  BinaryGAN \citep{dong2018training}, PathologyGAN \citep{quiros2019pathologygan}, and the retinal vessel dual-phase pipeline \citep{guibas2017synthetic} generate 28X28 pixels, 224X224 pixels, and 512X512 pixels size of images respectively. Inspired by pix2pixHD \citep{wang2018high}, we implemented three discriminators each one of them receiving a different scale of the input mask – 100\%, 50\%, and 25\% (Fig.~\ref{f:f2}B). This technique helps the discriminator to distinguish between real and fake masks from different levels of perspective. Low-resolution masks provide high-level information such as the general structure of the tissue. In contrast, low-level information, such as intercellular spaces, is obtained by high-resolution masks. The combined objective is provided by the following equation:
  
\begin{equation}
\mathcal{L}_{DEPAS}=\sum_{r=1}^{R} \alpha_{r} \cdot \mathcal{L}_{GAN,r} 
\label{eq_depas_loss}
\end{equation}

Where $\alpha_{r}$ is the weight of $D_r$, the discriminator that corresponds to the image resolution $r \in \{25\%, 50\%, 100\%\}$, and $R$ is the number of discriminators ($R=3$). $\mathcal{L}_{GAN,r}$ is the Vanilla GAN’s loss obtained by $D_{r}$ that receives the real input mask $x$ and the generator’s synthetic mask $G(z)$, and is defined as:
\begin{equation}
\mathcal{L}_{GAN,r} = log(D_{r}(x)) + log(1-D_{r}(G(z)))  \label{eq_gan_loss}
\end{equation}

\subsection{Paired Image Translation}
Paired image translation is a set of tasks that translate one domain of images to another domain given input-output image training pairs \citep{isola2017image}. One of these kinds of tasks is to insert a semantic map and translate it to an image, based on the additional information, such as class labels, passed together with the image to the network during the training phase. In the second step of our pipeline, we used pix2pixHD, an image translation generative network \citep{wang2018high}, to produce synthetic pathological images from the given semantic masks. Particularly, pix2pixHD consists of a generator which is a composition of convolutional residual layers that receives a 512X1024 pixels semantic mask as an input and generates a 512X1024 pixels RGB image. In addition, we used two multiscale discriminators with the same CNN architecture, but work on two different image scales. 

\subsection{Datasets}
In this study, we perform our methodology on four histology realizations subjected to two different types of staining.
The first type is hematoxylin and eosin (H\&E) where histology images were collected from three different types of cancer: Prostate Adenocarcinoma (PRAD), Skin Cutaneous Melanoma (SKCM), and Lung Squamous Cell Carcinoma (LUSC). The three datasets are part of the Cancer Genome Atlas (TCGA) research network, where for all datatypes we used only their imaging information \citep{tomczak2015review}. We performed our methodology on 50 WSIs from each H\&E realization. 

The second staining type is immunohistochemistry (IHC), where histology images were collected from non-small cell lung carcinoma (NSCLC) patients. This data was originally part of a study that involved an Immune Checkpoint inhibitor therapy, where the detection of programmed death-ligand 1 (PD-L1) in the tissue biopsies was required for determining the course of therapy \citep{wang2016pd}. $27$ WSIs were obtained from patients diagnosed with NSCLC who underwent biopsy at Rambam Health Care Campus. All procedures performed in this study and involving human participants were in accordance with the ethical standards of the Rambam Medical center institutional research committee,  approval 0522-10-RMB, and with the 1964 Helsinki declaration and its later amendments or comparable ethical standard.

In all realizations, the slides were split into patches with a size of 512X1024 pixels. Patches containing more than 85\% background were filtered. In total, 6000 images were used from each one of the H\&E datasets and 2012 images from the IHC dataset. For each realization, 85\% of the data was used for training and the rest for evaluation. In all realizations, we create ground truth for binary tissue semantic masks by converting the patches to grayscale and applying a high threshold to extract air pixels. We found that 204 and 235 were the optimal thresholds (in the range of 0-255) to distinguish between tissue and air pixels, in H\&E and IHC respectively.

\subsection{Training Procedure}
For each realization, we trained two models individually: pix2pixHD and DEPAS.
PixPixHD was trained with pairs of images and their corresponding tissue masks, where the tissue masks served as the input for the generator and the images as the real ground truth for the discriminators. The training was conducted with a batch size of 1 for 400 epochs in the H\&E realizations, and for 700 epochs in the IHC realization.  We linearly decay the learning rate to zero over the last 100 epochs for H\&E and over 200 epochs for IHC.

DEPAS was trained only with the tissue masks from the same training set used to train pix2pixHD. In this case, the generator generates synthetic tissue masks from a sample drawn from the standard normal distribution. The model was trained with a batch size of 8, for 100 epochs where every 10 epochs, the Annealing Sigmoid’s $\delta$ parameter in (\ref{eq_sig}) increased by one. Each scale objective in the loss term (\ref{eq_depas_loss}) was equally weighted where $\alpha_{100\%} = \alpha_{50\%} = \alpha_{25\%} = 1$. 
Both DEPAS and Pix2pixHD models' weights were optimized by Adam optimizer \citep{kingma2014adam} with a learning rate of 2e-4, and beta’s coefficients range of (0.5, 0.999). They were developed in the PyTorch framework \citep{paszke2019pytorch} and were trained on a single NVIDIA RTX A6000 GPU with 48GB GPU memory.

\section{Results}
\subsection{Synthetic Semantic Tissue Masks}
\begin{figure}[p]
\centering
	\includegraphics[scale=0.4]{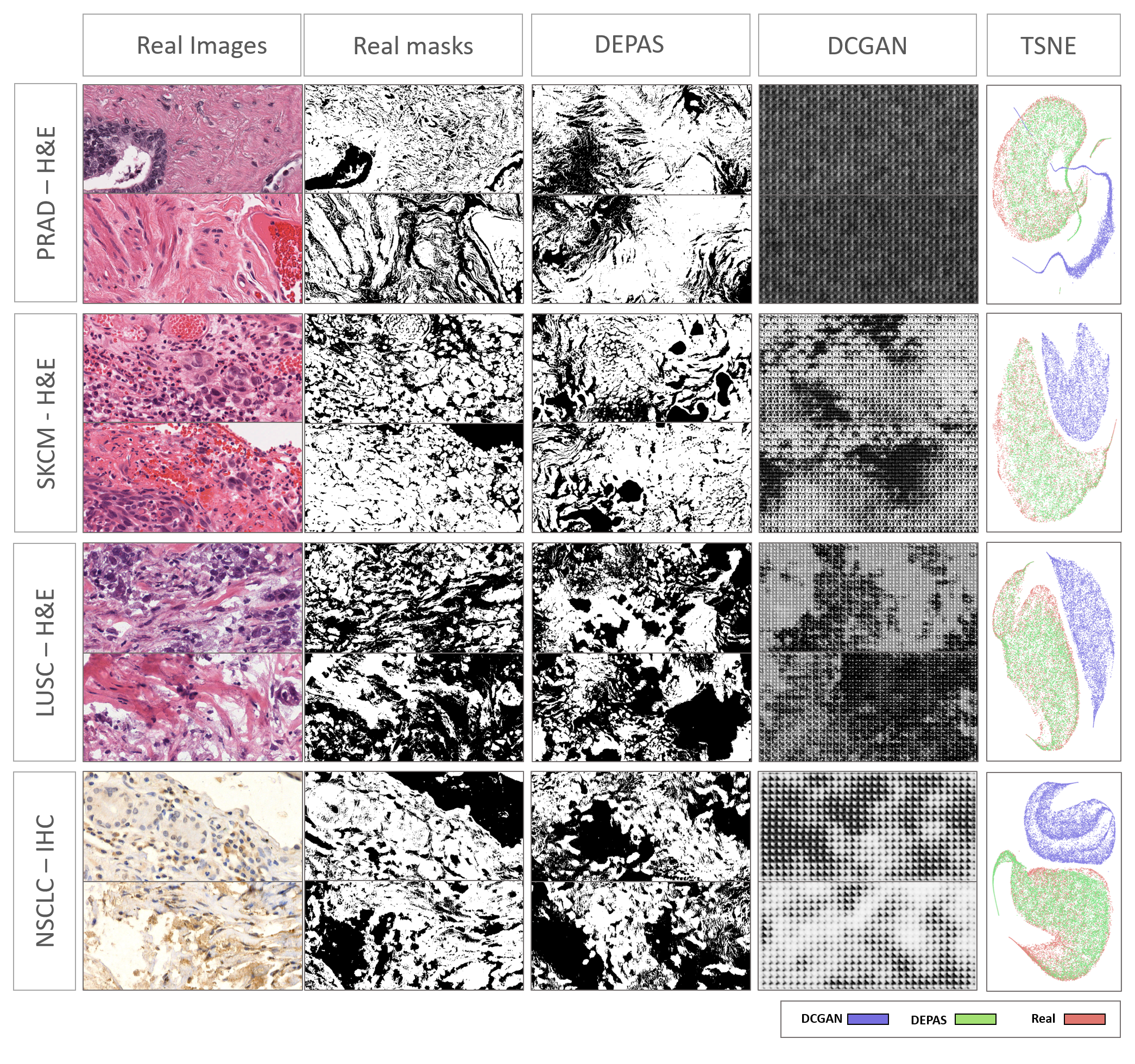}
	\caption{Examples of tissue masks from four different types of cancer realizations. These realizations include three organs: skin, prostate and lung, and two types of staining: H\&E and PD-L1 IHC. SKCM - Skin Cutaneous Melanoma. PRAD - Prostate Adenocarcinoma. LUSC - Lung Squamous Cell Carcinoma. NSCLC - Non-small cell lung carcinoma. In each realization, we show tissue masks taken from real biopsy slides (with their original RGB image). We compared the tissue masks to the ones produced by DEPAS and by a baseline DCGAN. The different types of tissue mask representations are projected into 2D via TSNE (right). We show that DEPAS provides tissue masks from a distribution that is closer to the Real images rather than DCGAN’s outputs (as quantified in Table I).}
	\label{f:f3}
\end{figure}

For each realization, we trained in addition to DEPAS, a standard DCGAN as a baseline since it was used before to generate semantic masks in the medical field \citep{guibas2017synthetic}. We generated from both trained models the same amount of synthetic tissue masks as in the real test set (1000 for H\&E realizations and 328 for the IHC). Examples of the different tissue masks for all four histology realizations are shown in Fig.~\ref{f:f3} with their latent space 2D projection extracted from a pre-trained ResNet model. 

For quantitative evaluation, we calculated the distance between DEPAS synthetic tissue masks to the real test tissue masks (that were excluded from training). The first two distance metrics, Kolmogorov–Smirnov (KS) test and Kullback–Leibler (KL) divergence, were applied to the TSNE projection of the masks' representations. These representations were extracted from the last hidden-layer of a pre-trained ResNet \citep{he2016deep} model (size of 2048). For the third metric, we captured representation vectors from the last layer of a pre-trained Inception v3 model \citep{szegedy2016rethinking} (size of 2048), to calculate ‘Fréchet inception distance’ (FID) which is the gold-standard metric for evaluating the quality of synthetic images \citep{heusel2017gans}.

The results for the different representation metrics are presented in Table \ref{tissue_mask_tab}. They show that for all four realizations, DEPAS synthetic tissue masks have the least distance to the real ones compared to DCGAN, in all metrics. Particularly, FID scores of DEPAS were lower (closer to the real images) than the DCGAN baseline by factors of 20.0, 6.9, 30.2,  and 17.0, when referring to the PRAD, SKCM, LUSC, and NSCLC tissue masks respectively.

\begin{table*}[p]
        \centering
	\begin{center}
		\small
    		\begin{tabular}{l c c c c c}
			\hline
		  & & \multicolumn{3}{c}{Image quality metrics}\\
            \cmidrule{3-5}
             Method & Dataset & KS $\downarrow$  & KL  $\downarrow$  & FID $\downarrow$  \\
			%\midrule
			\hline
			\hline
			DEPAS & \multirow{2}{*}{PRAD$^{a}$} & 1 (0.3) & 1 (0.9) &  1 (420.3) \\
   		DCGAN & & x2.6 & x42.8 & x20.0 \\
			\hline
            \hline
			DEPAS & \multirow{2}{*}{SKCM$^{a}$} & 1 (0.3) & 1 (0.7) & 1 (1006.6) \\
   		DCGAN & & x5.0 & x31.3 & x6.9  \\
        	\hline
            \hline
			DEPAS & \multirow{2}{*}{LUSC$^{a}$} & 1 (0.2) & 1 (0.6) & 1 (151.3) \\
   		DCGAN & & x11.5 & x67.7 & x30.2 \\
        	\hline
            \hline
			DEPAS & \multirow{2}{*}{NSCLC$^{b}$} & 1 (0.2) & 1 (0.3) & 1 (480.0) \\
   		DCGAN & & x6.6 & x128.0 & x17.0 \\
            \hline
		\end{tabular}
	\end{center}
	\caption{Similarity metrics between DEPAS synthetics masks, real masks, and current SOTA. $^a$ and $^b$ denote H\&E and IHC stained tissues respectively. In addition, KS is Kolmogorov–Smirnov test, KL is Kullback–Leibler divergence, and FID is Fréchet inception distance. Values are normalized by DEPAS results, where the parenthesis phrase presents its actual raw data. Downarrow symbol indicates lower is better.}
	\label{tissue_mask_tab}
\end{table*}

\begin{figure}[p]
\centering
         \includegraphics[ width=\textwidth]{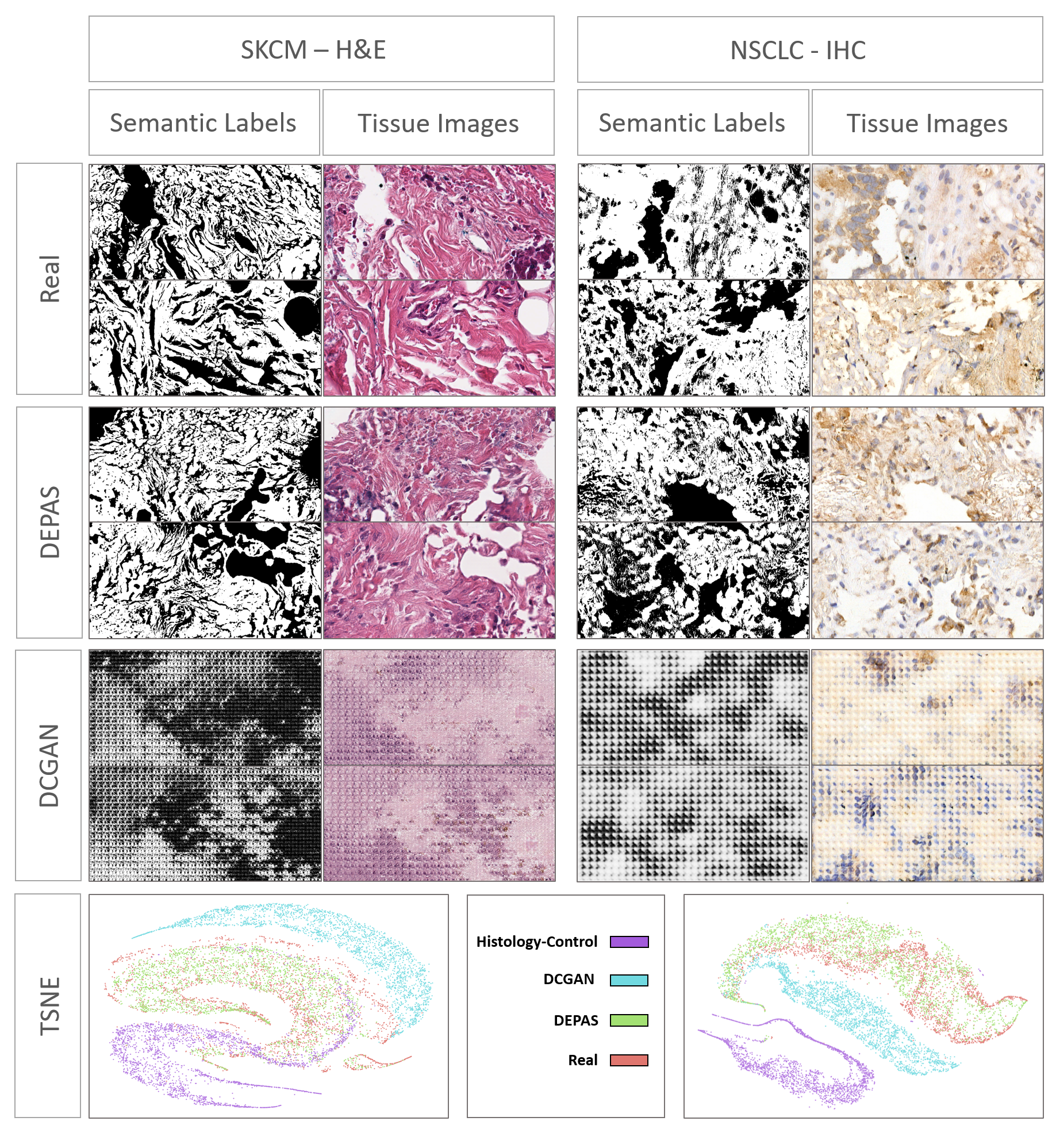}
	\caption{Examples of tissue masks and their corresponding histology RGB images for two types of cancers. H\&E stating of skin cutaneous melanoma (left) and IHC staining of non-small cell lung carcinoma (right). For each realization, we show pairs of masks-images taken from real biopsies slides. We compare them to the pairs produced by DEPAS and by a standard DCGAN. The different types of RGB image representations are projected into 2D via TSNE (bottom). We show that DEPAS provides images from a distribution that is closer to the Real images’ distribution rather than DCGAN’s outputs, or the negative histology-control outputs taken from real histology RGB images from a different realization.}
	\label{f:f4}
\end{figure}

\begin{table*}[t]
    \centering
	\begin{center}
		\small
    		\begin{tabular}{l c c c c c}
			%\toprule
			\hline
		  & & \multicolumn{3}{c}{Image quality metrics}\\
            \cmidrule{3-5}
             Method & Dataset & KS $\downarrow$  & KL $\downarrow$  & FID $\downarrow$  \\
			%\midrule
			\hline
			\hline
			DEPAS & \multirow{4}{*}{SKCM$^{a}$} & 1 (0.3) & 1 (0.4) & 1 (592.5) \\
   		  DCGAN & & x2.3 & x16.5 & x6.4  \\
                Pathology Control & & x1.8 & x8.8 & x9.5  \\
                Realistic Control & & x1.1 & x9.2 & x10.0  \\
        	\hline
                \hline
			DEPAS & \multirow{4}{*}{NSCLC$^{b}$} & 1 (0.2) & 1 (0.4) & 1 (219.9) \\
   		  DCGAN & & x3.0 & x8.2 & x1.4 \\
                Pathology Control & & x3.7 & x44.8 & x25.7  \\
                Realistic Control & & x2.8 & x111.8 & x40.2  \\
               \hline
		\end{tabular}
	\end{center}
	\caption{Similarity metrics between synthetic images based on DEPAS synthetics masks, real histological images, and various baselines. $^a$ and $^b$ denote H\&E and IHC stained tissues respectively. In addition, KS is Kolmogorov–Smirnov test, KL is Kullback–Leibler divergence, and FID is Fréchet inception distance. Values are normalized by DEPAS results, where the parenthesis phrase presents its actual raw data. The down arrow symbol indicates lower is better.}
	\label{rgb_tab}
\end{table*}

\subsection{Synthetic Photorealistic RGB Images}
To further evaluate the full pipeline in the photorealistic histology perspective, we applied the synthetic tissue masks to the image translation model and compared their outputs to the real histology images. We performed this over two datasets, the first is SKCM which represents a realization subjected to H\&E staining, and the second is lung cancer (NSCLC) subjected to IHC staining. Examples of the different tissue masks and images for both H\&E and IHC realizations are shown in Fig.~\ref{f:f4} with their latent space 2D projection extracted from a pre-trained ResNet model. 
Furthermore, we performed the same quantitative evaluation methodology over the images at the RGB level, by calculating the distance between the synthetic images stem from DEPAS to the real RGB histology images (Table \ref{rgb_tab}). For comparison, we performed the same for three other datasets: (1) synthetic images where their prior tissue masks are generated by DCGAN. (2) Real histology images from a different type of pathology as a histology control. I.e. in the case where the evaluation is performed on the H\&E realization, we also calculated the distance between the real IHC images to the real H\&E images as a histology baseline, and vice-versa. (3) We also used SegTrack Dataset \citep{li2013video} as real-life scenario images for realistic control taken from the non-pathology field. 

We performed the same evaluation for all control batches in RGB levels. The results in Table \ref{rgb_tab} show that for both H\&E and IHC images, DEPAS had the best performance over all metrics. Where DEPAS’s FID score for H\&E images was better than DCGAN by a factor of 6.36, and by a factor of  1.42 for IHC images.

\subsection{Multi-Label DEPAS}
We further show the ability of DEPAS to generate synthetic multi-label discrete semantic masks on the IHC realization. This task is performed on the NSCLC dataset as before, but where the tissue mask is divided into more-detailed labels, based also on its PD-L1 attributes. PD-L1 is a molecule expressed by tumor cells and enables them to evade the immune system’s attack. Hence, a common immunotherapy treatment uses blocking antibodies that target PD-L1 to increase the immune system's effectiveness against the tumor cells. Evaluating the PD-L1 rate in the patients’ biopsies is essential for determining the treatment type and its level. To achieve PD-L1-related labels, all IHC patches were annotated by expert pathologists. For every patch, each tissue pixel was assigned to one of the four PD-L1 feature classes: Inflammation, PD-L1- and PD-L1+, or non of them. Two more classes were assigned using computer vision techniques. Air class was assigned as before where grayscale pixels with values higher than 235 were considered Air. The cells class was assigned to the darker values of brown. Empirically, they were captured in the RGB image representation where the green and blue channels were smaller than 200 and the red channel was higher than the other channels. 

\begin{figure}[thpb]
\centering
	\includegraphics[scale=0.47]{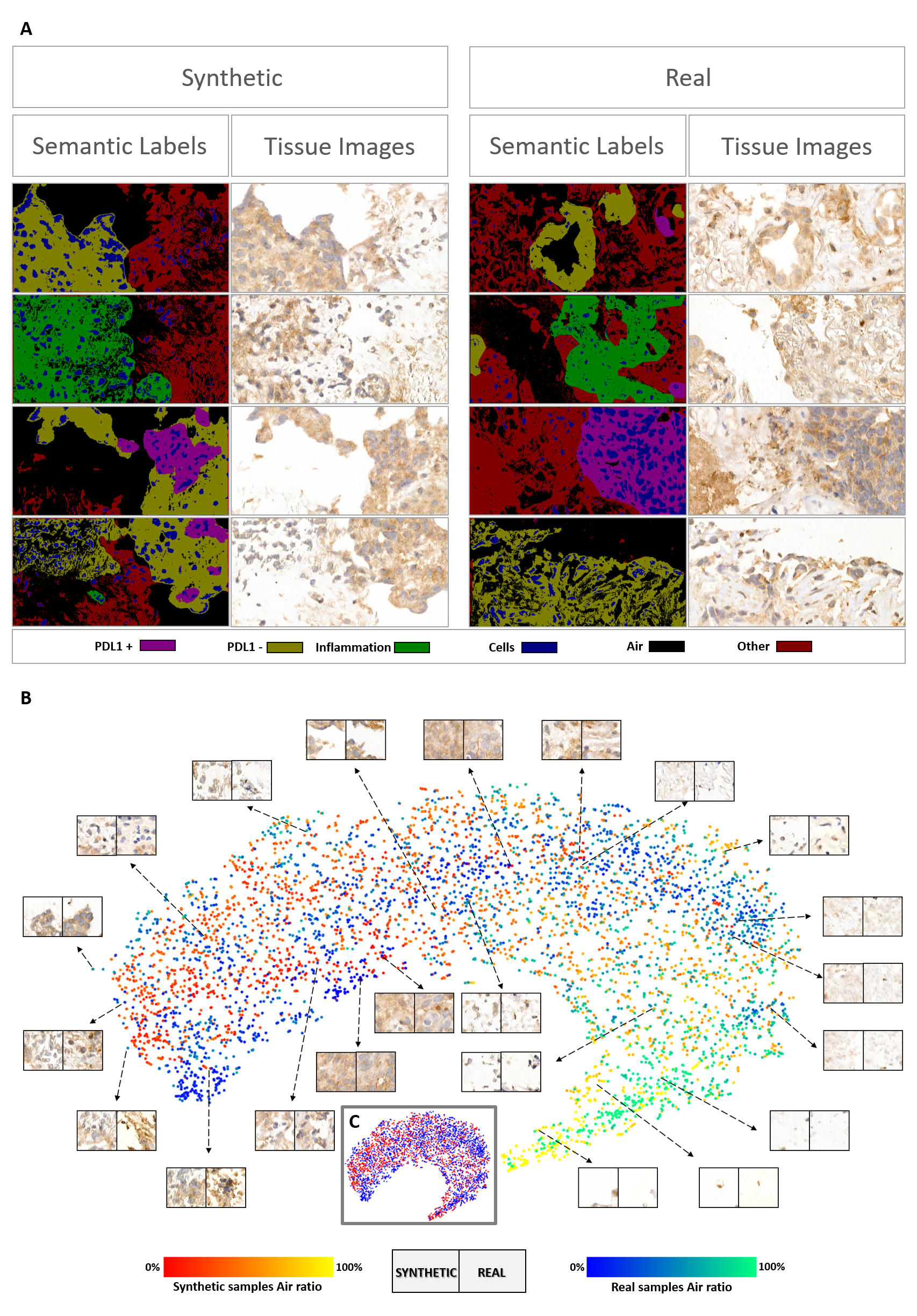}
	\caption{(A) Examples of a multi-label task that consists of PD-L1 tissue attributes as semantic labels. DEPAS’s synthetic labels and images are shown alongside real PD-L1 examples as a reference. (B) presents a TSNE projection of inception’s representations taken from both real and synthetic images’ (after being cropped into 224X224 pixels). Autumn and Winter colormaps represent synthetic and real images projection respectively. As the colormaps are brighter, the histology images contain more air. Several pairs of real images and synthetic images that had the smallest Euclidian distance between them are shown as well. (C) displays the same TSNE projection but with a single coloring of the data types to emphasize the mixture between the synthetic (red) and real (blue).
}
	\label{f:f5}
\end{figure}

After creating the multi-label Ground truth, we trained the image translation model and DEPAS using the same methodology and hyperparameters, with only one exception. In the multi-label case, we train DEPAS by adjusting the discrete adaptive block to produce a multi-channel feature map. In this case, we set the initial value of the temperature in (\ref{eq_temp}) to 1, and divide it every 10 epochs by 1.25. Examples of synthetic semantic masks generated from DEPAS, and their corresponding synthetic histology images, are presented in Fig.~\ref{f:f5}A and visualized near real examples. We projected the images' inception representation via TSNE into a 2D space and show that the real images get mixed up with the synthetic ones. We also present representing pairs of real and synthetic images that had the smallest Euclidian distance in the 2D space (Fig.~\ref{f:f5}B). Furthermore, synthetic RGB images generated by the multi-label approach are closer to the real images than the ones generated in the binary approach by 14\% when referring to FID scores.

\section{CONCLUSION}
One of the main challenges of generating synthetic images of tissues is controlling the distribution of features within them. Paired GANs provide a good way to improve the synthetics image quality by introducing semantic masks that account for the spatial structure of the tissue. However,  Unlike other domains, such as autonomous vehicles or face recognition, simulating or generating synthetic masks that represent the actual biological complexity of the image with high fidelity are lacking.
A reasonable compromise, is augmenting semantic information taken from real tissues to generate photorealistic histology images. Yet, this approach suffers from bounded scalability, when a limited amount of real data are required for inference.

Here we introduce an architecture for generating high-resolution binary masks of tissue structure that can be used as semantic prior knowledge for image translation models. Our work copes with the main challenge of generating binary synthetic images by adding noise along the different stages of decoding blocks and adding annealing temperature blocks to overcome the undifferentiability that is associated with binary masks while processing by a differential pipeline through the generator and discriminator. Our approach allows us not only to generate synthetic binary masks but also to produce multilabel masks that are critical for many applications that require labeling different cellular regions (such as cancer cells). 

We show that our synthetic mask indeed captures the real cell distribution and spatial orientation within various pathological realizations histology slides.  Moreover, we show the synthetic images that result from our masks resemble real histological images better than the baseline in two types of cancers subjected to H\&E and IHC staining types. Furthermore, we show that performing our pipeline with more detailed tissue information reflected in the multi-label semantic masks improves the quality of the synthetic images.

Overall, our work provides a state-of-the-art solution for the challenging task of synthetic histological image generation with their semantic information in a form that is both scalable and controllable.
\\
\\

\section*{ACKNOWLEDGMENT}
The authors would like to thank Tanya Wasserman for her technical support and valuable discussions. The results shown here are in part based upon data generated by the TCGA Research Network: https://www.cancer.gov/tcga.

\bibliographystyle{plainnat}

\begin{thebibliography}{}

\bibitem{czyzewski2021machine}
T.~Czyzewski, N.~Daniel, M.~Rochman, J.~M. Caldwell, G.~A. Osswald, M.~H.
  Collins, M.~E. Rothenberg, and Y.~Savir, ``Machine learning approach for
  biopsy-based identification of eosinophilic esophagitis reveals importance of
  global features,'' {\em IEEE open journal of engineering in medicine and
  biology}, vol.~2, pp.~218--223, 2021.

\bibitem{daniel2022deep}
N.~Daniel, A.~Larey, E.~Aknin, G.~A. Osswald, J.~M. Caldwell, M.~Rochman, M.~H.
  Collins, G.-Y. Yang, N.~C. Arva, K.~E. Capocelli, {\em et~al.}, ``A deep
  multi-label segmentation network for eosinophilic esophagitis whole slide
  biopsy diagnostics,'' in {\em 2022 44th Annual International Conference of
  the IEEE Engineering in Medicine \& Biology Society (EMBC)}, pp.~3211--3217,
  IEEE, 2022.

\bibitem{larey2022fron}
A.~Larey, E.~Aknin, N.~Daniel, G.~A. Osswald, J.~M. Caldwell, M.~Rochman,
  T.~Wasserman, M.~H. Collins, N.~C. Arva, G.-Y. Yang, M.~E. Rothenberg, and
  Y.~Savir, ``Harnessing artificial intelligence to infer novel spatial
  biomarkers for the diagnosis of eosinophilic esophagitis,'' {\em Frontiers in
  Medicine}, vol.~9, 2022.

\bibitem{goodfellow2020generative}
I.~Goodfellow, J.~Pouget-Abadie, M.~Mirza, B.~Xu, D.~Warde-Farley, S.~Ozair,
  A.~Courville, and Y.~Bengio, ``Generative adversarial networks,'' {\em
  Communications of the ACM}, vol.~63, no.~11, pp.~139--144, 2020.

\bibitem{mirza2014conditional}
M.~Mirza and S.~Osindero, ``Conditional generative adversarial nets,'' {\em
  arXiv preprint arXiv:1411.1784}, 2014.

\bibitem{isola2017image}
P.~Isola, J.-Y. Zhu, T.~Zhou, and A.~A. Efros, ``Image-to-image translation
  with conditional adversarial networks,'' in {\em Proceedings of the IEEE
  conference on computer vision and pattern recognition}, pp.~1125--1134, 2017.

\bibitem{wang2018high}
T.-C. Wang, M.-Y. Liu, J.-Y. Zhu, A.~Tao, J.~Kautz, and B.~Catanzaro,
  ``High-resolution image synthesis and semantic manipulation with conditional
  gans,'' in {\em Proceedings of the IEEE conference on computer vision and
  pattern recognition}, pp.~8798--8807, 2018.

\bibitem{adjei2020gan}
P.~E. Adjei, Z.~M. Lonseko, and N.~Rao, ``Gan-based synthetic gastrointestinal
  image generation,'' in {\em 2020 17th International Computer Conference on
  Wavelet Active Media Technology and Information Processing (ICCWAMTIP)},
  pp.~338--342, IEEE, 2020.

\bibitem{armanious2020medgan}
K.~Armanious, C.~Jiang, M.~Fischer, T.~K{\"u}stner, T.~Hepp, K.~Nikolaou,
  S.~Gatidis, and B.~Yang, ``Medgan: Medical image translation using gans,''
  {\em Computerized medical imaging and graphics}, vol.~79, p.~101684, 2020.

\bibitem{lysik2022he}
M.~Lysik, Z.~Swiderska-Chadaj, T.~Markiewicz, T.~Les, S.~Cierniak, and
  M.~Lorent, ``He-to-pas histological stain conversion by gan in renal
  pathology,'' in {\em 2022 International Joint Conference on Neural Networks
  (IJCNN)}, pp.~1--7, IEEE, 2022.

\bibitem{ghahremani2022deepliif}
P.~Ghahremani, J.~Marino, R.~Dodds, and S.~Nadeem, ``Deepliif: An online
  platform for quantification of clinical pathology slides,'' in {\em
  Proceedings of the IEEE/CVF Conference on Computer Vision and Pattern
  Recognition}, pp.~21399--21405, 2022.

\bibitem{divya2022medical}
S.~Divya, L.~P. Suresh, and A.~John, ``Medical mr image synthesis using
  dcgan,'' in {\em 2022 First International Conference on Electrical,
  Electronics, Information and Communication Technologies (ICEEICT)},
  pp.~01--04, IEEE, 2022.

\bibitem{smaida2021dcgan}
M.~Smaida, S.~Yaroshchak, and Y.~El~Barg, ``Dcgan for enhancing eye diseases
  classification.,'' in {\em CMIS}, pp.~22--33, 2021.

\bibitem{puttagunta2022novel}
M.~Puttagunta, R.~Subban, {\em et~al.}, ``A novel covid-19 detection model
  based on dcgan and deep transfer learning,'' {\em Procedia computer science},
  vol.~204, pp.~65--72, 2022.

\bibitem{blanco2021medical}
R.~F. Blanco, P.~Rosado, E.~Vegas, and F.~Reverter, ``Medical image editing in
  the latent space of generative adversarial networks,'' {\em
  Intelligence-Based Medicine}, vol.~5, p.~100040, 2021.

\bibitem{quiros2019pathologygan}
A.~C. Quiros, R.~Murray-Smith, and K.~Yuan, ``Pathologygan: Learning deep
  representations of cancer tissue,'' {\em arXiv preprint arXiv:1907.02644},
  2019.

\bibitem{guibas2017synthetic}
J.~T. Guibas, T.~S. Virdi, and P.~S. Li, ``Synthetic medical images from dual
  generative adversarial networks,'' {\em arXiv preprint arXiv:1709.01872},
  2017.

\bibitem{neff2017generative}
T.~Neff, C.~Payer, D.~Stern, and M.~Urschler, ``Generative adversarial network
  based synthesis for supervised medical image segmentation,'' in {\em Proc.
  OAGM and ARW joint Workshop}, vol.~3, p.~4, 2017.

\bibitem{bengio2013estimating}
Y.~Bengio, ``Estimating or propagating gradients through stochastic neurons,''
  {\em arXiv preprint arXiv:1305.2982}, 2013.

\bibitem{dong2018training}
H.-W. Dong and Y.-H. Yang, ``Training generative adversarial networks with
  binary neurons by end-to-end backpropagation,'' {\em arXiv preprint
  arXiv:1810.04714}, 2018.

\bibitem{chung2016hierarchical}
J.~Chung, S.~Ahn, and Y.~Bengio, ``Hierarchical multiscale recurrent neural
  networks,'' {\em arXiv preprint arXiv:1609.01704}, 2016.

\bibitem{tomczak2015review}
K.~Tomczak, P.~Czerwi{\'n}ska, and M.~Wiznerowicz, ``Review the cancer genome
  atlas (tcga): an immeasurable source of knowledge,'' {\em Contemporary
  Oncology/Wsp{\'o}{\l}czesna Onkologia}, vol.~2015, no.~1, pp.~68--77, 2015.

\bibitem{wang2016pd}
X.~Wang, F.~Teng, L.~Kong, and J.~Yu, ``Pd-l1 expression in human cancers and
  its association with clinical outcomes,'' {\em OncoTargets and therapy},
  vol.~9, p.~5023, 2016.

\bibitem{kingma2014adam}
D.~P. Kingma and J.~Ba, ``Adam: A method for stochastic optimization,'' {\em
  arXiv preprint arXiv:1412.6980}, 2014.

\bibitem{paszke2019pytorch}
A.~Paszke, S.~Gross, F.~Massa, A.~Lerer, J.~Bradbury, G.~Chanan, T.~Killeen,
  Z.~Lin, N.~Gimelshein, L.~Antiga, {\em et~al.}, ``Pytorch: An imperative
  style, high-performance deep learning library,'' {\em Advances in neural
  information processing systems}, vol.~32, 2019.

\bibitem{he2016deep}
K.~He, X.~Zhang, S.~Ren, and J.~Sun, ``Deep residual learning for image
  recognition,'' in {\em Proceedings of the IEEE conference on computer vision
  and pattern recognition}, pp.~770--778, 2016.

\bibitem{szegedy2016rethinking}
C.~Szegedy, V.~Vanhoucke, S.~Ioffe, J.~Shlens, and Z.~Wojna, ``Rethinking the
  inception architecture for computer vision,'' in {\em Proceedings of the IEEE
  conference on computer vision and pattern recognition}, pp.~2818--2826, 2016.

\bibitem{heusel2017gans}
M.~Heusel, H.~Ramsauer, T.~Unterthiner, B.~Nessler, and S.~Hochreiter, ``Gans
  trained by a two time-scale update rule converge to a local nash
  equilibrium,'' {\em Advances in neural information processing systems},
  vol.~30, 2017.

\bibitem{li2013video}
F.~Li, T.~Kim, A.~Humayun, D.~Tsai, and J.~M. Rehg, ``Video segmentation by
  tracking many figure-ground segments,'' in {\em Proceedings of the IEEE
  international conference on computer vision}, pp.~2192--2199, 2013.

\end{thebibliography}

\end{document}